\documentclass[preprint]{aastex}
\begin{document}
\title{Strong Lensing Reconstruction}
\author{Ue-Li Pen}
\affil{Canadian Institute for Theoretical Astrophysics,
60 St George St., Toronto, Ont. M5S 3H8, Canada }

\newcommand{\etal}{{\it et al. }}
\newcommand{\beq}{\begin{equation}}
\newcommand{\eeq}{\end{equation}}

\begin{abstract}

We present a general linear algorithm for measuring the surface mass
density $1-\kappa$ from the observable reduced shear
$g=\gamma/(1-\kappa)$ in the strong lensing regime.  We show that in
general, the observed polarization field can be decomposed into
``electric'' and ``magnetic'' components, which have independent and
redundant solutions, but orthogonal noise properties.  By
combining these solutions, one can increase the signal-to-noise ratio
by $\sqrt{2}$.  The solutions allow dynamic optimization of signal and
noise, both in real and Fourier space (using arbitrary smoothing
windows).  Boundary conditions have no effect on the reconstructions,
apart from its effect on the signal-to-noise.  Many existing
reconstruction techniques are recovered as special cases of this
framework.  The magnetic solution has the added benefit of yielding the
global and local parity of the reconstruction in a single step.

\end{abstract}

\keywords{(cosmology:) gravitational lensing}

\newpage

Gravitational lensing is rapidly providing large, independent data sets
which are direct measures of gravitational potentials.  One would like
to understand the properties of gravitational lensing reconstruction to
optimize the solutions in a similar fashion as has been done for galaxy
surveys (Vogeley and Szalay 1996).  In the weak lensing regime, the
reconstruction is linear, and optimization of signal-to-noise becomes a
straightforward problem.  When lensing becomes strong, the equations
appear to become non-linear, and the solutions are more difficult to
understand.  Of particular interest is the understanding of error
propagation, to quantify the confidence of the solution, and to
optimize the signal-to-noise for real, noisy data sets.  The
measurement of galaxy polarization, or reduced shear, yields two
observable quantities.  The ellipticity $e=(1-R)/(1+R)$ is defined in
terms of the minor/major axis ratio $R$, and the orientation of galaxies
$\varphi$ is measured relative to the x-axis at
each pixel of the reconstruction.  One constructs a spin-2 vector with
components $e_1=e \cos 2\varphi,\ e_2=e\sin 2\varphi$.  The reduced
shear $g_i=e_i$ in the even parity case (i.e. for weak lensing), but
changes to $g_i=e_i/e^2$ across a critical line.
In the single source-lens plane
lensing problem, the only unknown is the dimensionless surface mass
density $\kappa$.  One would thus expect, in general, two independent
solutions to exist, each using only half the observable information.
If the noise properties in each solution are understood, and
orthogonal, the two solutions can be combined in an optimal fashion.
It is the purpose of this paper to systematically construct such
solutions.  We also quantify the existing solution algorithms in the
same framework, and discuss error properties.



We assume single source redshifts, which is a reasonable approximation
for low redshift lenses.  
We will denote vectors and matrices by either
bold symbols or explicitly using indices.  One proceeds as follows:  Let
\newcommand{\bG}{{\bf \Gamma}}
\beq
\bG \equiv \left( \begin{array}{cc}
		\gamma_1 & \gamma_2 \\
		\gamma_2 & -\gamma_1
	\end{array} \right)
\eeq
\newcommand{\bg}{{\bf G}}
and the observable reduced shear is given as $\bg=\bG/(1-\kappa)$,
i.e. $(1-\kappa)g_a=\gamma_a$.
There are two observables $g_1,g_2$, but only one
independent unknown $\kappa$.

Let us first solve for $1-\kappa$ by noting that $\bG$ has no magnetic
component (Seljak 1997), i.e. 
\beq
\Gamma_{lm}=(2\partial_l\partial_m\nabla^{-2}-\delta_{lm}) 
\partial_i\partial_j \nabla^{-2} \Gamma_{ij}
\eeq
which we can recast in terms of the observable reduced shear
\beq
(1-\kappa)G_{lm}=E_{lm}\equiv (2\partial_l\partial_m\nabla^{-2}-\delta_{lm}) 
\partial_i\partial_j \nabla^{-2} (1-\kappa)G_{ij}.
\label{eqn:kappa}
\eeq
We note that this is a linear differential equation for $1-\kappa$ in
terms of the reduced shear.  It is formally a fourth order equation.
A peculiarity of equation
(\ref{eqn:kappa}) is that it is both overdetermined, and singular.  In
other words, $1-\kappa$ is the null eigenvector of the magnetic
equation
\beq
(1-\kappa)B_{lm}\equiv (1-\kappa)G_{lm}-E_{lm}=0.
\label{eqn:blm}
\eeq
\newcommand{\bx}{{\bf x}}
\newcommand{\bk}{{\bf k}}
In the presence of noise, equation (\ref{eqn:kappa}) may not have any
solutions.  We can choose instead a linear least squares problem where
we minimize the noise $S^B$:
\beq
S^B\equiv \int B_{ij} B^{ij} d^2{\bf x}.
\label{eqn:min}
\eeq
$S^B$ is a quadratic function in $1-\kappa$, which we wish to minimize.
Because of the global invariance transformation (Falco \etal 1985,
Seitz and Schneider 
1995), we cannot determine $1-\kappa$ up to a multiplicative constant,
and in fact $S^B$ would naively be minimized by $\kappa=1$.   We
impose a Lagrange multiplier L=$\int (1-\kappa)^2 d^2{\bf x} - 1=0$,
and minimize $S^B+\lambda L$ for $1-\kappa$ and $\lambda$
respectively.  The first variation leads to a linear eigenvector
problem of the form ${\bf A}(1-\kappa)=\lambda(1-\kappa)$, while the
The linear matrix ${\bf A}$ can be written as
\begin{eqnarray}
A^B_{\alpha\beta} &\equiv& \frac{\partial^2 S^B}{\partial \kappa({\bf
x}_\alpha) \partial \kappa(\bx_{\beta})}
\\
&=& \left(\sum_{i,j}G_{ij}(\bx_\alpha)^2\right)\delta(\bx_\alpha-\bx_\beta)
 - 2G_{ij}(\bx_\alpha)G_{lm}(\bx_\beta) \int
e^{i \bk \cdot (\bx_\alpha-\bx_\beta)} \hat{k}_i \hat{k}_j 
\hat{k}_l\hat{k}_m  d^2\bk
\label{eqn:a}
\end{eqnarray}
where $\hat{k}^i=k_i/|k|$.
In real space, the integral becomes
\begin{eqnarray}
A^B_{\alpha\beta}&=&\frac{-2}{|\bx_\alpha-\bx_\beta|^2} 
G_{ij}(\bx_\alpha) G_{lm}(\bx_\beta)
\left[ \delta_{il}\delta_{jm}
-4\hat{x}^i\hat{x}^l\delta_{jm}
+ 4\hat{x}^i\hat{x}^j\hat{x}^l\hat{x}^m \right]
\nonumber \\
&&+\frac{1}{2}
	G(\bx_\alpha)^2\delta(\bx_\alpha-\bx_\beta)
\nonumber \\
&=&e^2(\bx_\alpha)\delta(\bx_\alpha-\bx_\beta)  
 +\frac{4}{|\bx_\alpha-\bx_\beta|^2} 
\bigg\{
\left[e_1(\bx_\alpha)e_1(\bx_\beta)+e_2(\bx_\alpha)e_2(\bx_\beta)\right]
\nonumber\\
&&-2[e_1(\bx_\alpha)\cos(2\theta)-e_2(\bx_\alpha)\sin(2\theta)]
[e_1(\bx_\beta)\cos(2\theta)-e_2(\bx_\beta)\sin(2\theta)]\bigg\}
\label{eqn:ab}
\end{eqnarray}
where
$\hat{\bx}=(\bx_\alpha-\bx_\beta)/|\bx_\alpha-\bx_\beta|=\{\cos(\theta),
\sin(\theta)\}$ and $G=\sqrt{\sum_{ij}G_{ij}^2}$.  

${\bf A}^B$ is formally infinite on the diagonal, but we can rescale
it onto a new matrix which is zero on the diagonal:
\beq
M^B_{\alpha\beta} \equiv \frac{A^B_{\alpha\beta}}{G(\bx_\alpha)
G(\bx_\beta)}-\frac{\delta(\bx_\alpha-\bx_\beta)}{2}.
\label{eqn:m}
\eeq
We note that the smallest eigenvector $v$ of ${\bf M}$ is
given by $v=(1-\kappa)G$.  One should note that across a critical
line, the observable $\bg$ changes parity, and one observes $\bg/G^2$
instead (see for example Kaiser 1995, herafter K95).  One of the main
features of this method is its continuity 
across critical lines.  We simply write down both solutions,
\begin{eqnarray}
\kappa_1&=&1-v/G
\nonumber \\
\kappa_2&=&1-v/G^3.
\label{eqn:v}
\end{eqnarray}
In a reasonably large field of view, where one knows the edges
to be outside
the outer critical line, one uses $\kappa_1$.  At the point where
$\kappa_1=\kappa_2$, we know that we have encountered a critical
curve, and we switch the solution variable to $\kappa_2$.  We will
need to do the same across the inner critical curve.  The local parity
ambiguity can now be solved after the global solution, which is one of
its main attractions.  We will discuss a local deterministic procedure
to determine the parity of (\ref{eqn:v}) below.

A direct solution for the smallest eigenvector of ${\bf M}$ can be
computationally expensive.  For an image which is $N$ pixels on each
side, ${\cal O}(N^6)$ operations are required for direct matrix
solvers.  This is the same operation count as directly solving the
Kaiser and Squires (1993) (hereafter KS) procedure without the use of
a fast Fourier transform (FFT), which becomes necessary if
non-periodic boundary conditions are imposed.  Linear algebra is,
fortunately, a well exploited subject, and highly optimized and
parallel libraries are available to find the required eigenvectors for
matrices as big as $2^{16}$ on a side in a day using a modern 100 CPU
multiprocessor, which is sufficient to directly reconstruct images with
$256^2$ pixels.  In any case, it would clearly be desirable to use an
iterative method, where each iteration would only involve a
convolution.  Since we are interested in the smallest eigenvector, an
inverse power method would yield rapid convergence.  Each such
iteration involves solving a linear system, which is again
straightforward using an iteration, since we know ${\bf A}$ to be
positive semi-definite.  The actual iterations are in fact just
convolutions, which could also be accelerated using FFTs.  The
convergence rate for the inverse power method is given by the ratio of
second smallest eigenvalue to smallest eigenvalue
$\lambda_2/\lambda_1$, which approaches infinite speed for small
noise. 

The standard ``electric'' mode reconstruction can be applied in a
similar fashion, but requires prior knowledge of the local parity.  
Let $u=1-\kappa$, $H_{ij}=G_{ij}+\delta_{ij}$, then
the lensing equation reads (in the even parity case)
\beq
\partial_i\partial_j u H_{ij}=0
\label{eqn:e}
\eeq
for which we can again define a least squares action $S^E=\int 
(\nabla^{-2}\partial_i\partial_j u H_{ij})^2$, resulting in a matrix
\begin{eqnarray}
A^E_{\alpha\beta}&=&\frac{1}{r^2}\left\{-2\hat{x}^i\hat{x}^j
[G_{ij}(\bx_\alpha)+
G_{ij}(\bx_\beta)]+G_{ij}(\bx_\alpha) G_{lm}(\bx_\beta)
\left[ \delta_{il}\delta_{jm}
-4\hat{x}^i\hat{x}^l\delta_{jm}
+ 4\hat{x}^i\hat{x}^j\hat{x}^l\hat{x}^m \right] \right\}
\nonumber \\
&&+\delta(\bx_\alpha-\bx_\beta)\left[1+\frac{G^2(\bx_\alpha)}{4}\right]
\label{eqn:ae}
\end{eqnarray}
where $r=|\bx_\alpha-\bx_\beta|$.
We note the following property of minimum eigenvector solutions to
(\ref{eqn:ae}): if $u$ is a solution for a given $H_{ij}$, then it is
also a solution for addition of magnetic noise, $H_{ij}+N^B_{ij}/u$.
We are considering noise fields ${\bf N}$ which are arbitrary
traceless tensor fields, which can be decomposed into electric and
magnetic components, in analogy with Equations
(\ref{eqn:kappa},\ref{eqn:blm}). 
This is the opposite property of the magnetic solution for
(\ref{eqn:m}), where one could add  electric noise to
$G_{ij}+N_{ij}^E/v$ and keep any solution $v$ invariant.  We see that
the two solutions 
have orthogonal dependencies on the noise, and thus expect their
combination to improve signal-to-noise ratios by $\sqrt{2}$.
If the noise is known to contribute equally to E and B, as most
sources would be expected to, we can use the difference between the
two solutions as an estimate of the noise.

An elegant observation by Kaiser (K95) was the
realization that $\partial_j u 
H_{ij}$ is a curl-free vector for any true solution $u$, allowing an
integration of (\ref{eqn:e}):
\beq
\partial_k\log(u)=-H^{-1}_{ki}\partial_jH_{ij}.
\label{eqn:k95}
\eeq
We can also solve an equivalent least squares problem by setting $S^K=\int
\nabla^{-2}(\partial_j u H_{ij})(\partial_k u H_{ik}) d^2\bx$.  This
strategy was explored by Lombardi and Bertin (1999), who considered
accelerated direct solutions.
The matrix is
\begin{eqnarray}
A^K_{\alpha\beta}&=&\frac{-2}{r^2}\left\{\hat{x}^i\hat{x}^j[G_{ij}(\bx_\alpha)+
G_{ij}(\bx_\beta)+G_{il}(\bx_\alpha)G_{lj}(\bx_\beta)]
\right\}
\nonumber\\
&&+\delta(\bx_\alpha-\bx_\beta)\left[1+\frac{G(\bx_\alpha)^2}{2}\right].
\label{eqn:ak}
\end{eqnarray}
\newcommand{\curl}{\nabla\times}
One also obtains ${\bf A}^K$ by a contraction of the appropriate
term in ${\bf A}^E$ (\ref{eqn:ae}).  
As K95 pointed out, the parity can be directly
determined from the curl of a vector.  
Let us define $\nu_i=\partial_j
u G_{ij}$.  We can, in analogy to the
magnetic solution, solve for both the parity and $\kappa$ by requiring
that $\curl {\bf \nu}=0$.
The corresponding action is $S^C=\int (\nabla^{-2} \curl {\bf
\nu})^2$, which results in a linear combination of the electric and
Kaiser matrices, ${\bf A}^C={\bf A}^K-{\bf A}^E$.  K95 proposed a
similar procedure to determine the parity, which is necessary to
construct equation (\ref{eqn:k95}) from the observable ellipticities $e_i$.

It is instructive to examine the weak lensing limit.  When $G\ll 1$,
the eigenvector of $H_{ij}$ can be considered as a perturbation on the
background $H^0_{ij}=\delta_{ij}$ with $H^1_{ij}=G_{ij}$.  We note
that Equations 
(\ref{eqn:ae}) and (\ref{eqn:ak}) agree to ${\cal O}(G)$.  We will
choose a Fourier weight function $k^2$ in the action, so $S^0=\int
(\partial_i u)\partial^iu$, and $S^1=\int
\nabla^{-2}(\partial_i\partial_j uG_{ij})^2$.
This breaks the degeneracy of
eigenvectors in ${\bf A}^0$, giving us $A^0=-\delta''(r)$.  
The eigenvectors
are the Fourier modes $\exp(i\bk\cdot\bx)$ with eigenvalue $k^{2}$.  The
zero eigenvector is given by $u^0=1$.  The perturbed lowest eigenvalue
is $\int k_i k_j G_{ij}$ (see for example Schiff 1971, section 31).
The perturbed zero 
eigenvector is given in Fourier space as the matrix element
\beq
u^1(\bk_\alpha) = \int \exp(-i\bk_\alpha\cdot\bx_\alpha)
\frac{A^1_{\alpha\beta}}{k^2}
d^2\bx_\beta d^2\bx_\alpha
\eeq
We then have $u^1=k_ik_jG_{ij}/k^2$, which is
exactly the KS solution.  The B equation (\ref{eqn:ab}) does not have
a weak lensing limit, since all entries are ${\cal O}(G^2)$.

One can combine the algorithms by adding the actions, for example
using $S=S^E+S^B$ or any linear combination of the three actions
described above.  The resulting matrix ${\bf A}$ just becomes the sum
of the matrices.  An interesting combination is $S^G=S^B+2S^E$, which has
no quadratic terms in ${\bf G}$ except on the diagonal.  
We note in passing that this $S^G=\sum_{ij}(S_{ij}^G)^2$ is in fact the least
squares action, or 
likelihood, of the appropriately weighted reconstructed reduced shear
as used in ``maximum likelihood reconstruction'' (Bartelmann \etal 1996):
\beq
S^G_{ij}=\nabla^{-2}\left(\partial_i\partial_j
-\frac{\delta_{ij}}{2}\nabla^2\right)\kappa 
-(1-\kappa)G_{ij}=0.
\label{eqn:g}
\eeq
At each point, $S^G/(1-\kappa)^2$ is just the difference squared
between the observed reduced shear $G_{ij}$ and the reconstruction for
a given $\kappa$ field.

The
construction of $S^B$ requires no prior knowledge of parity, while
$S^E$ does, so one might expect to proceed by first solving $S^B$ and
then using the weak lensing equation directly.
The B-type solution (\ref{eqn:m}) has only made use of the ${\bf B}$
component of the reduced shear, which works well in the strong lensing
regime, but results in very poor optical depth estimation in the weak
lensing regime.  One can apply a linear method, such as KS on the
reconstructed (unreduced) shear 
\beq
\bG^o=\frac{v\bg}{G}.  
\label{eqn:bg}
\eeq
No parity knowledge is needed for equation (\ref{eqn:bg}),
which automatically returns the correct parity for $\bG$.  We can then
reconstruct $\kappa^{\rm KS}$ from $\bG$, and compare this second
solution with (\ref{eqn:v}).  This allows us to locally and
globally determine the parity of the solution.  In the final map, one
can combine the KS inferred $\kappa^{\rm KS}$ with the $\kappa^o$
inferred using the new procedure.

The construction of ${\bf M}$ from ${\bf A}$ can be generalized to
optimize for prior knowledge of noise and signal statistics.  Equation
(\ref{eqn:min}) can be generalized with arbitrary filter weightings in
Fourier space,
\beq
S\equiv \int B_{ij}(\bk) B^{ij}(-\bk) W(|k|) d^2{\bk}.
\eeq
The window function $W(k)$ transforms to $W(r)$ in real space.
The matrix entries are accordingly modified as follows:
\begin{eqnarray}
A^B_{\alpha\beta}&=&G_{ij}(\bx_\alpha)G_{ij}(\bx_\beta)\left[
W(r)+2V_1(r)-2V_2(r)\right]
\nonumber\\
&&-2G_{ij}(\bx_\alpha) G_{lm}(\bx_\beta)
\bigg\{
4\hat{x}^i\hat{x}^l\delta_{jm}\left[
6V_1(r)-3V_2(r)+V_3(r)\right]
\nonumber \\
&&+\hat{x}^i\hat{x}^j\hat{x}^l\hat{x}^m\left[-15V_1(r)+15 V_2(r)
-6V_3(r)+V_4(r)\right]
\bigg\}
\nonumber\\
A^E_{\alpha\beta}&=& W(r)
+ [G_{ij}(\bx_\alpha)+
G_{ij}(\bx_\beta)]\hat{x}^i\hat{x}^j\left[W(r)-2W_2(r)\right]
\nonumber \\
&&+G_{ij}(\bx_\alpha) G_{lm}(\bx_\beta)
\bigg\{
4\hat{x}^i\hat{x}^l\delta_{jm}\left[
6V_1(r)-3V_2(r)+V_3(r)\right]
\nonumber \\
&&+\hat{x}^i\hat{x}^j\hat{x}^l\hat{x}^m\left[-15V_1(r)+15 V_2(r)
-6V_3(r)+V_4(r)\right]
\bigg\}
\nonumber \\
A^K_{\alpha\beta}&=&W(r)
+ \left[G_{ij}(\bx_\alpha)+
G_{ij}(\bx_\beta)+G_{ik}(\bx_\alpha)G_{kj}(\bx_\beta)\right]
\nonumber\\
&&\times
\left\{\delta_{ij}W_2(r)+\hat{x}^i\hat{x}^j\left[W(r)-2W_2(r)\right]
\right\}
\end{eqnarray}
where $V_n(r)=r^{4-n}\partial^n_r \nabla^{-4}W(r)$ and
$W_n(r)=r^{-n}\int^r s^{n-1}W(s)ds$.  The original formulae
(\ref{eqn:ab},\ref{eqn:ae},\ref{eqn:ak})
are reproduced for $W(r)=\delta(r)$ (2-dimensional).
In addition, we can impose a weighting in real space, as done in
Equation (\ref{eqn:m}).  Instead of dividing to obtain a unit
diagonal, one might also envision weighting by the expected local
signal-to-noise (Pen 1999).  In general, we see that setting
$M_{\alpha\beta}=A_{\alpha\beta} w(\bx_\alpha)w(\bx_\beta)$ relates
the zero eigenvector $u(\bx_\alpha)$ of $A_{\alpha\beta}$ to the
computed zero eigenvector $m(\bx_\alpha)$ of $M_{\alpha\beta}$ by
$u(\bx_\alpha)=m(\bx_\alpha)w(\bx_\alpha)$.  We see immediately that
no boundary condition artifacts are ever introduced when one truncates
${\bf A}$ at arbitrary boundaries, for example by choosing $w$ to be
1 on the domain with data and 0 elsewhere.
One should note that if too much noise is added to $S$, it can happen
that the smallest eigenvector is dominated by noise, while the second
smallest eigenvector actually contains the correct solution.

We have presented a direct linear solution strategy for the strong
lensing problem.  The B-type solutions work contiguously across
critical lines and critical density lines without prior parity
knowledge.  The critical lines are then obtained from the solutions
(\ref{eqn:v}) themselves.  The existence and uniqueness of the
solutions and the critical lines are therefore clearly established,
which is in general not clear for non-linear solution schemes.  Its
noise properties depend only on the B-type noise, in contrast to the
KS weak lensing and the E-type reconstruction procedure which depends
only on E-type noise.  We have fully exploited the fact that the two
observables $g_1,g_2$ allow the construction of two independent
solutions with orthogonal noise properties.  Combining them explicitly
allows one to improve the signal-to-noise by up to $\sqrt{2}$.  In
addition, it allows for arbitrary weightings in both real and Fourier
space.  We have shown how to apply the same procedure to the K95 and
standard maximum likelihood algorithms.  The solutions furthermore
allow arbitrary weightings in both real and Fourier space, which
allows optimization of signal-to-noise in the final reconstruction.

I would like to thank Robert Schmidt, Roman Scoccimarro, Albert
Stebbins, Peter Schneider and Lindsay King for helpful discussions.
The Max-Planck Institut f\"ur Astrophysik provided the wonderful
hospitality where the work was completed during the GAAC workshop.
The work was supported in part by NSERC grant 72013704 and NASA grant
NAG5-7039.

\end{document}